\documentstyle[preprint,aps]{revtex}
\tightenlines

\begin{document}

\title{Multi-frequency evaporative cooling to BEC
in a high magnetic field.}

\author{V. Boyer, S. Murdoch, Y. Le Coq, G. Delannoy, P. Bouyer, A. Aspect}

\address{Groupe d'Optique Atomique Laboratoire Charles Fabry
de l'Institut d'Optique,\\ UMRA 8501 du CNRS,\\ B\^{a}t. 503, Campus
universitaire d'Orsay,\\ B.P. 147, F-91403 ORSAY CEDEX, FRANCE}

\maketitle

\begin{abstract}
We demonstrate a way to circumvent the interruption of evaporative
cooling observed at high bias field for $^{87}$Rb atoms trapped in
the $(F=2, m=+2)$ ground state. Our scheme uses a
3-frequencies-RF-knife achieved by mixing two RF frequencies. This
compensates part of the non linearity of the Zeeman effect,
allowing us to achieve BEC where standard 1-frequency-RF-knife
evaporation method did not work. We are able to get efficient
evaporative cooling, provided that the residual detuning between
the transition and the RF frequencies is smaller
than the power broadening of the RF transitions at the end of the
evaporation ramp.

\end{abstract}


Forced evaporative cooling of atoms \cite{KetterleEVAP,forcedevap} in a
magnetic trap is at the moment the only known way to achieve
Bose-Einstein condensation \cite{CornellBEC,HuletBEC,KetterleBEC}.
Particles with energy significantly larger than the average
thermal energy are removed from the trap and the remaining ones
thermalize to a lower temperature by elastic collisions. For that,
a radiofrequency (RF) magnetic field is used to induce a multi-photon
transition from a trapping state to a non-trapping state via all
intermediate Zeeman sublevels. Atoms moving in the trap with
sufficient energy can reach the resonance point (RF knife) and
exit the trap. If the RF-frequency is decreased slowly enough, and
no other process is hampering the forced-evaporation, the increase
of the phase space density obtained by this method eventually
leads to Bose-Einstein condensation.

In a previous publication \cite{Des99}, we reported that RF forced
evaporative cooling of $^{87}$Rb atoms in the $(F=2,m=+2)$ ground
state in a magnetic trap with a high bias field is hindered and
eventually interrupted. Our interpretation of this phenomenon is
based on the non-linear terms of the Zeeman effect that lift the
degeneracy of transition frequencies between adjacent Zeeman
sublevels. This interpretation is supported by numerical
calculations \cite{suo99}. Interrupted evaporative cooling in a
large magnetic field is a serious problem in several situations,
interesting for practical reasons \verb|-| like the use of permanent
magnets \cite{Tollett} or of an iron core electromagnet as the one
described in \cite{Des98}. High magnetic field evaporation is also
important in connection with Feshbach resonances
\cite{Ketterle,Chu,Heinzen,Wieman}. In this paper, we demonstrate
that it is possible to achieve efficient evaporative cooling in a
high magnetic field, by use of a multi-frequency RF knife allowing
a multi-photon transition to take place across non equidistant
levels. We show that, for our range of magnetic fields, it is
possible to use a simple experimental scheme where the three
required frequencies are obtained by RF frequency mixing yielding
a carrier and two sidebands.



We focus in this paper on $^{87}$Rb in the $F=2$ manifold of the
electronic ground state. Atoms are initially trapped in the $m=+2$
state. 
Our high bias field magnetic trap follows the Ioffe-Pritchard
scheme. To the second order in position (see eq. 1 in \cite{Des99}), 
the magnetic field modulus 
$B$ has a 3D
quadratic dependence allowing trapping, plus a bias field $B_{0}$
 between 50 and 200 Gauss. This is
much larger than in most other experiments where $B_{0}$ can be
independently adjusted, and is set typically at 1 Gauss
\cite{TOP}. In a large magnetic field, the non linear terms are
not negligible in the Zeeman shifts given by the Breit-Rabi
formula
\begin{equation}
E_{m}(B)=mg_{{\rm I}}\mu _{{\rm n}}B+\frac{\hbar
\omega _{{\rm HF}}}{2}\left( \sqrt{1+m\xi +\xi ^{2}}-1\right)
\label{Breit_Rabi}
\end{equation}
with
\begin{displaymath}
\xi =\frac{(g_{{\rm S}}\mu _{{\rm B}}+g_{{\rm I}}\mu _{{\rm n}})B}{\hbar
\omega _{{\rm HF}}}.
\end{displaymath}
Here $g_{{\rm S}}\simeq 2.002$ and $g_{{\rm I}}\simeq 1$ are
respectively the Land\'{e} factor for the electron and the nucleus,
$\mu _{B}$ and $\mu _{n}$ are the Bohr magneton and the nucleus
magneton, and $\omega _{{\rm HF}}$ ($2\pi \times 6834.7$ MHz) is
the hyperfine splitting.

Compared to the low magnetic field case
\cite{KetterleEVAP,forcedevap}, the evaporation process changes
drastically. At a given magnetic field, the spacings between
adjacent sublevels ($|\Delta m|=1$) are not equal and the direct
multi-photon transition from trapping to non-trapping states
becomes negligible. Evaporation of hot atoms can only happen via a
sequence of one-photon transitions of limited efficiency (see fig.
8 in \cite{Erice}) separated in space. This results in long
lasting atoms in the $m=+1$ and $m=0$ states \cite{m0} responsible
for hindered evaporative cooling. Moreover, transitions to
non-trapping states are suppressed at the end of the evaporation
ramp, leading to an interruption of cooling before BEC is reached.

To overcome these limitations, 3 distinct RF fields can be used to
induce a direct three photon transition from the $m=+2$ trapping
state to the $m=-1$ non trapping state. At a magnetic field $B$,
the three RF frequencies must match the transition frequencies
defined by:
\begin{eqnarray}
\omega _{0}-\delta \omega _{0}^{\prime } &=&(E_{2}-E_{1})
/\hbar \nonumber \\
\omega _{0} &=&(E_{1}-E_{0})/\hbar\label{3KnivesValues}
  \\
\omega _{0}+\delta \omega _{0} &=&(E_{0}-E_{-1})/\hbar   \nonumber
\end{eqnarray}
with $E_{m}$ taken from eq.(\ref{Breit_Rabi}).

Fig. \ref{3knives} represents all possible transitions induced by
these three RF frequencies in the magnetic trap. At position $K$, each
RF field is resonant with a given transition : the smallest RF
frequency with the $(m=+2)\rightarrow (m=+1)$ transition, the
intermediate frequency with the $(m=+1)\rightarrow (m=0)$
transition, and the largest frequency with the $(m=0)\rightarrow
(m=-1)$ transition; this is where the 3-photon transition occurs.
Because of the ordering of the three RF frequencies, the points
where one-photon transitions can be induced from $m=+2$ to $m=+1$
by the two larger frequencies are located beyond $K$ (the
multi-photon knife). Consequently, during the evaporation, hot
atoms will first encounter the three photon knife and be expelled
from the trap, provided that the RF power is large enough to
enable efficient multi-photon adiabatic passage to the
non-trapping state $m=-1$.


The discussion above shows that in principle, the multi-frequency
evaporation requires a synchronized non trivial sweep of three
different frequencies in the 100 MHz range, with an accuracy of a
few kHz (see below). We have rather implemented a simplified
scheme where the three frequencies are obtained by mixing a
carrier at frequency $\omega_{\rm RF}$ with a smaller frequency
$\delta \omega_{\rm RF} $. We then obtain three equally spaced
radiofrequency fields : $\omega_{\rm RF} - \delta\omega_{\rm RF}$,
$\omega_{\rm RF}$, $\omega_{\rm RF} + \delta\omega_{\rm RF}$, of
approximately the same power (as checked with a spectrum
analyzer). Since in general $\delta \omega _{0}$ and $\delta
\omega _{0}^{\prime }$ are slightly different, the RF frequencies
will not exactly match the transition frequencies of
eq.(\ref{3KnivesValues}). Nonetheless, they compensate the second
order (quadratic) term of the Zeeman shift, and should work under
certain condition discussed hereafter.

At the position where the three-photon transition is resonant,
the carrier frequency $\omega_{\rm RF}$ will verify
\begin{equation}
3\omega _{0}+\delta \omega _{0}-\delta \omega _{0}^{\prime
}=3\omega_{\rm RF} \label{res3cout}
\end{equation}
but there will be a residual detuning for each one photon step of
the multi-photon transition. For example, the optimum
$\delta\omega_{\rm RF}$ that maximizes the multi-photon transition
probability will be
\begin{equation}
\delta \omega_{\rm RF} =\frac{\delta \omega _{0}+\delta \omega _{0}^{\prime
}}{2} \label{deltaOpti}
\end{equation}
and the residual detunings for each intermediate steps of the
three photons transition are both equal to
\begin{equation}
\Delta =\frac{\delta \omega _{0}-\delta \omega _{0}^{\prime}}{6}.
\end{equation}
If the Rabi frequency $\Omega_{\rm RF}$ associated with each one
photon transition is significantly larger than the residual
detuning $\Delta$, the multi-photon transition is quasi resonant
in the intermediate levels, leading to an effective Rabi frequency
$\Omega_{\rm eff}\propto\Omega_{\rm RF}$. If on the other hand
$\Omega_{\rm RF}$ is smaller than $\Delta$, the effective Rabi
frequency is
\begin{equation}
\Omega_{\rm eff}\propto\frac{\Omega_{\rm RF}^{3}}{\Delta^{2}}
\end{equation}
and the multi-photon transition is inefficient for evaporation ;
we are then in the scheme of hindered and interrupted evaporation.
We therefore expect that our scheme will be efficient for small
enough  magnetic field when the residual detuning $\Delta$ is
smaller than the one-photon Rabi frequency $\Omega_{\rm RF}$.

Table \ref{table1} gives the values of the Zeeman shifts and the difference 
$\delta\omega_{0}-\delta \omega_{0}^{\prime}$ for
various magnetic fields. For the RF power used in
this scheme, the one photon Rabi frequency $\Omega_{\rm RF}$ is of
the order of 10 kHz, and the discussion above shows that our
simplified 3-knives evaporation scheme should work for magnetic
fields significantly less than a hundred Gauss. This is what we
observe: it is impossible to achieve BEC in bias fields of 207
Gauss and 110 Gauss, but BEC is obtained in a trap with a bias
field of 56 Gauss, by using an appropriate sideband splitting
$\delta \omega_{\rm RF}$ kept constant while ramping down the
carrier frequency $\omega_{\rm RF}$.

Figure \ref{Rabi} shows the effect of the sideband splitting
$\delta\omega_{\rm RF} $ at a bias field value of 56 Gauss. We
have plotted the number of condensed atoms as a function of
$\delta\omega_{\rm RF}$, all other parameters being kept
unchanged. This is a good indication of the efficiency of the
evaporation. The curve shows a maximum at $\delta\omega_{\rm RF} =
2\pi\times 0.45$ MHz. This value verifies equation
(\ref{deltaOpti}) for a magnetic field of 56.6 Gauss. This
magnetic field corresponds to the position of the RF knife at the
end of the ramp. We conclude that frequency matching is mostly
important in the last part of the radiofrequency ramp. The width
of the curve is about 10 kHz (HWHM) which corresponds to power
broadening \cite{LZ}.

Table \ref{table2} report experimental data, showing
quantitatively the efficiency of our simplified 3-knives scheme,
without which BEC could not be obtained at 56 Gauss. It is
interesting to note that even when the magnetic field is too large
to allow our simplified 3-knives scheme to reach BEC, it is
nevertheless more efficient than a simple 1-frequency knife, since
it allows us to reach a significantly lower temperature. It is
also remarkable that an efficient evaporation was obtained at a
bias field of 56 Gauss, since the beginning of the evaporation
takes place in a larger magnetic field (of the order of 200 Gauss)
where the condition (\ref{deltaOpti}) does not hold, and the
detuning of the intermediate one photon transitions is much larger
than the Rabi frequency $\Omega_{\rm RF}$. Although it has not
been much noticed, a similar situation is encountered in most BEC
experiments (using 1-frequency knife evaporation) : the non linear
Zeeman effect at the beginning of the evaporation is often much
larger than the Rabi frequency, and the evaporation hampering
described in \cite{Des99} is certainly happening then. The success
of these experiments as well as of our 3-frequencies scheme shows
that whether the evaporation is hindered or not only matters at
the end of the evaporation ramp. To understand qualitatively this
observation, we can note that the heating induced by the atoms
populating the intermediate levels should not vary drastically
with the temperature of the cooled cloud. At the beginning of the
evaporation, i.e. ``high'' temperatures, the relative heating
stays negligible \cite{Jul97}. Close to the end, i.e. ``low''
temperature, when heating should give rise to hampered evaporative
cooling, evaporation is fully efficient and the intermediate
levels are completely depleted. This could explain the success of
BEC experiments. To verify these assumptions, more theoretical
work, for instance in the spirit of \cite{suo99}, is needed.

In conclusion, we have demonstrated a scheme to circumvent the
hindrance and interruption of evaporative cooling in the presence
of non linear Zeeman effect. We implement a 3-frequency
evaporative knife by a modulation of the RF field, yielding two
sidebands. This scheme allows us to obtain BEC of $^{87}$Rb atoms
in the $(F=2,m=+2)$ ground state in a bias field of 56 Gauss,
where the standard 1-frequency RF evaporation scheme fails. Our
observations also support the physical ideas presented in our
previous work to explain the hindrance and interruption of
evaporative cooling in a high magnetic field, as well as the
qualitative discussions of this paper.

The success of this simplified scheme and the complementary
observations reported in this paper, indicate that a more
sophisticated multi-frequency evaporation scheme should work at
larger bias field, provided that the resonance in the intermediate
steps of the multi-photon transition is achieved within the Rabi
frequency of the one photon transitions, at the end of the
evaporative ramp.

\begin{acknowledgments}
The authors thank S. Rangwala for helpfull discussions and M.
L\'ecrivain for the elaboration of the iron-core electromagnet.
This work is supported by CNRS, MENRT, R\'{e}gion Ile de France, DGA
and the European Community. SM acknowledges support from Minist\`{e}re
des Affaires \'{E}trang\`{e}res. YLC acknowledges support from DGA.
\end{acknowledgments}

\begin{figure}[p]
\caption{Implementation of 3-RF-knife to evaporate in a high
magnetic field. All possible transitions are represented.
Evaporation happens at $K$ via a 3-photon-transition resonant in
the intermediate states.} \label{3knives}
\end{figure}

\begin{figure}[p]
\caption{Bose Einstein condensation with 3 RF knives : number of
atoms in the condensate versus the sideband frequency $\delta
\omega_{\rm RF}$. The width of the curve is of the order of the
Rabi frequency of a one-photon RF transition.} \label{Rabi}
\end{figure}

\begin{table}[htb]
\begin{tabular}{c|c|c|c}
$B$ (Gauss) & 56 & 110 & 207 \\  \hline \hline
$\omega_{0}-\delta\omega_{0}'$ ($2\pi\times$ MHz) & 39.058-0.434 &
76.255-1.621 & 141.800-5.398
\\ \hline
$\omega_{0}$ ($2\pi\times$ MHz) & 39.058 & 76.255 & 141.800 \\
\hline $\omega_{0}+\delta\omega_{0}$ ($2\pi\times$ MHz) & 39.058+0.449 &
76.255+1.732 & 141.800+6.096 \\ \hline \hline
$\delta\omega_{0}-\delta\omega_{0}'$ ($2\pi\times$ kHz) & 15 & 111 & 698
\end{tabular}
\caption{Zeeman effect for different magnetic fields, calculated
from the Breit-Rabi formula. \label{table1}}
\end{table}

\begin{table}[htb]
\begin{tabular}{c|c|c|c}
$B_{0}$ (Gauss) & 56 & 110 & 207 \\ \hline  \hline
$T_{{\rm 1 knife}}$($\mu$K)
& 10 & 50 & 100 \\ \hline $T_{{\rm 3 knives}}$ ($\mu$K) & 0.1 &
0.5 & 15 \\ \hline $n\lambda^{3}_{\rm 3 knives}$ & $>$ 2.612 & 0.1
& $10^{-3}$
\end{tabular}
\caption{Experimental results : lowest temperature achievable with
and without side band activated, and highest phase space density
achieved for different bias field. At a bias field of 56 Gauss,
our 3-frequency scheme yields BEC, while a single frequency scheme
fails because of interrupted evaporative cooling. \label{table2}}
\end{table}


\begin{references}
\bibitem{TOP} Note however that in TOP traps
the bias field often exceeds 10 Gauss, and hindered cooling may
play a role, specially when the atomic ground state hyperfine
splitting is relatively small as in sodium for example.
\bibitem{m0}
The $m=0$ state is a trapping state in this manifold because of
the non linearity of the Zeeman effect.
\bibitem{LZ} This conclusion was corroborated by a calculation of
the energies of the dressed states for a given set $\{ \Omega_{\rm
RF}, \omega_{\rm RF}, \delta\omega_{\rm RF} \}$. From the
calculated energy splitting $C\simeq\Omega_{\rm eff}$ at the
$(m=+2,m=-1)$ level crossing, we used the two-levels Landau Zener
probability that the atoms will follow an adiabatic transition. We
verified that for small Rabi frequencies (i.e. small evaporation
efficiency), as in our experiment, this 3-photon transition is the
most probable transition at any sideband detuning
$\delta\omega_{\rm RF}$.  This numerical calculation can be used
to fit the experimental data. We can estimate the one-photon Rabi
frequency $\Omega_{\rm RF}=2\pi\times 8\pm 4$ kHz.
\end{references}
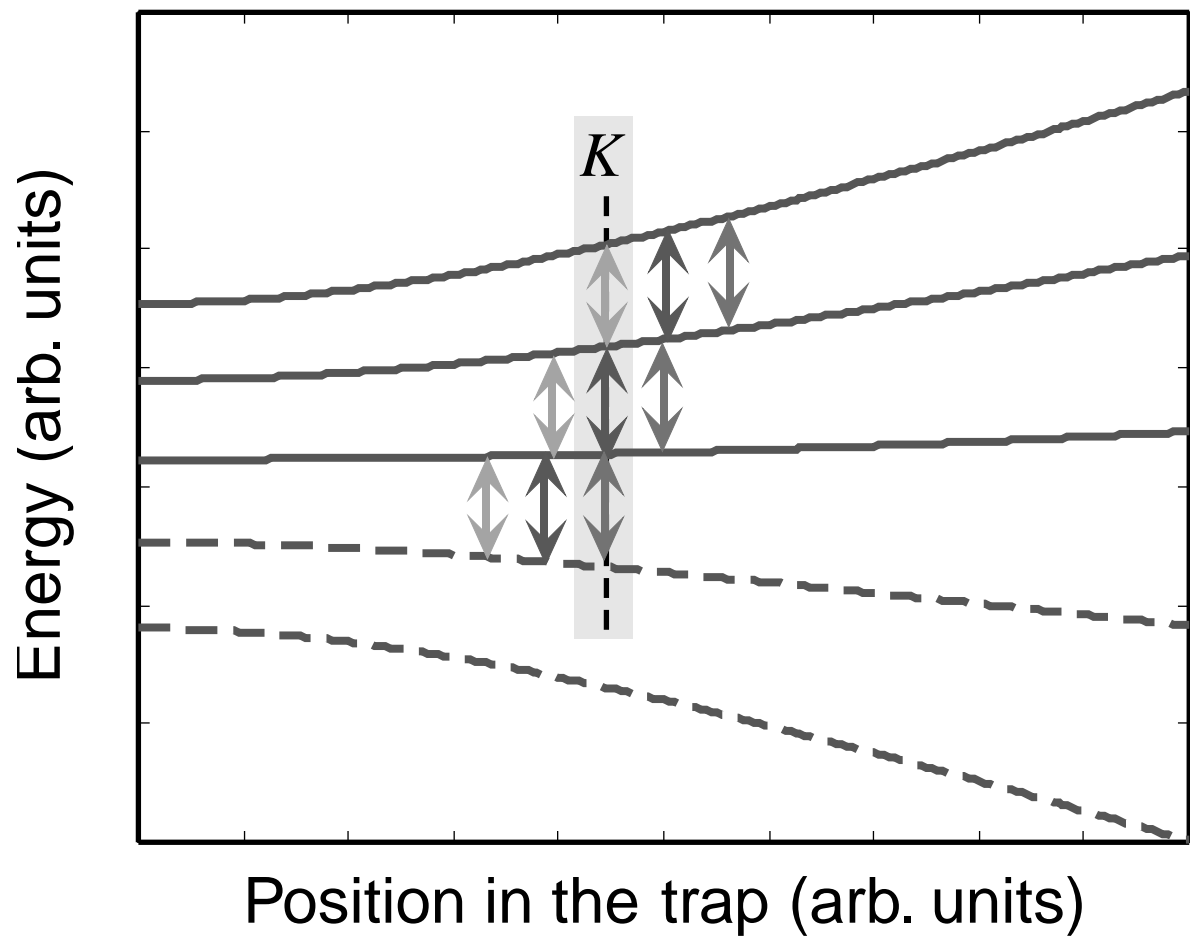
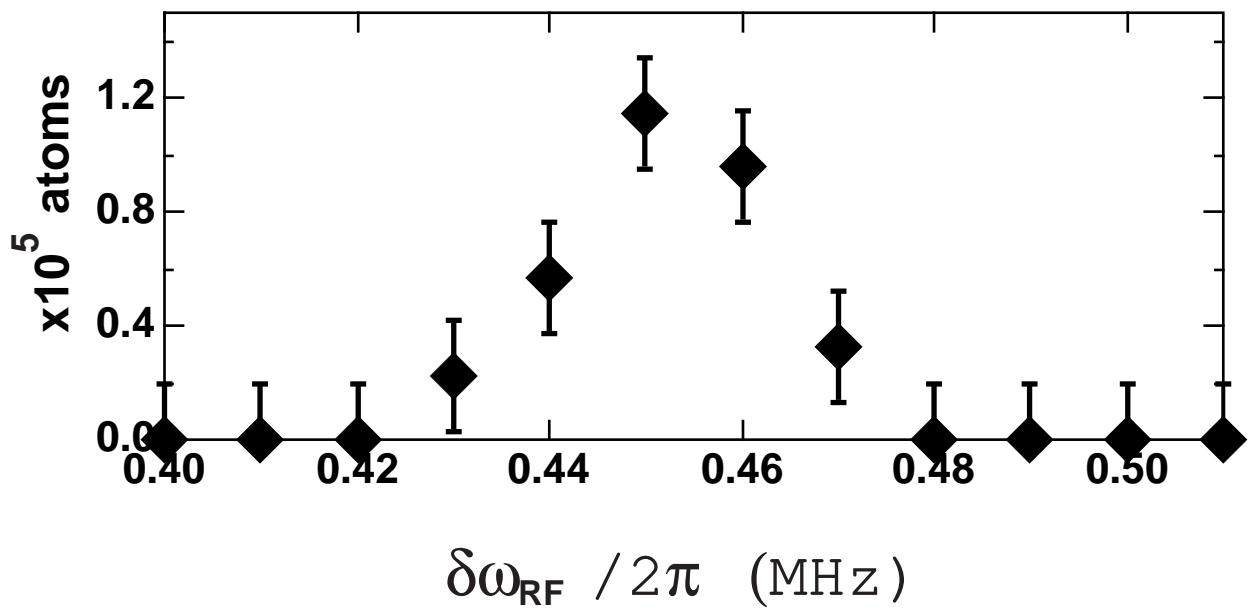
\end{document}